\documentclass{iaus}
\usepackage{graphicx}

\title[The Slow Winds of A supergiants] 
{The Slow Winds of A--Type Supergiants}

\author[Anah\'i Granada, Michel Cur\'e \& Lydia Cidale]   
{Anah\'i Granada$^{1,2,4}$,
 Michel Cur\'e$^3$ and Lydia Cidale$^{1,2}$}

\affiliation{$^1$Facultad de Ciencias Astron\'omicas y Geof\'isicas, Universidad Nacional de La Plata, Argentina\\
$^2$Instituto de Astrof\'isica La Plata, CCT La Plata-CONICET-UNLP, Argentina\\
$^3$Universidad de Valpara\'iso, Chile \\
$^4$Observatoire de Gen\`eve. Universit\'e de Gen\`eve, Suisse\\[\affilskip] 
email: {\tt granada@fcaglp.unlp.edu.ar}}

\pubyear{2010}
\volume{272}  
\pagerange{1--2}
\jname{Active OB stars: structure, evolution, mass loss and critical limits.}
\editors{C. Neiner, G. Wade, G. Meynet \& G. Peters, eds.}
\begin{document}

\maketitle

\begin{abstract}
The line driven- and rotation modulated-wind theory predicts an alternative 
{\it slow} solution, besides from the standard m-CAK solution, when the 
rotational velocity is close to the critical velocity. We study the behaviour 
of the winds of A-type supergiants (Asg) and show that under particular conditions, 
e.g., when the $\delta$ line-force 
parameter is about 0.25, the slow solution 
could exist over the whole star, even for the cases when the rotational speed 
is slow or zero. We discuss density and velocity profiles as well as possible 
observational conterparts.
\end{abstract}
\firstsection
\section{Introduction}
The theory of radiation driven winds (CAK, \cite[Castor et al. 1975]{CAK75}) 
and later m-CAK (\cite[Pauldrach et al. 1986]{Pau86}) succeeded in 
 describing terminal velocities (V$_{\infty}$) and mass losses ($\dot{M}$) of hot stars apart from
predicting the wind momentum luminosity (WML) relationship. 
 This relationship was empirically found for the most of 
luminous O-type stars (\cite[Puls et al. 1996]{Pul96}) 
and  extended to lower 
luminosity objects by \cite[Kudritzki et al. (1999)]{Kud99}, who found that it depends on  the spectral type.  
 Particularly, Asg show V$_\infty$ values a factor
3 lower than theoretical ones 
(\cite[Achmad et al. 1997]{Ach97}) and 
V$_{\infty}$ decreases when increasing the escape velocity (V$_{esc}$) (\cite[Verdugo et al. 1999]{Ver99}), in clear
 contradiction with the CAK theory (fast solution). Moreover, \cite[Kudritzki et al. (1999)]{Kud99}
 found that H$\alpha$ profile of these stars can be modeled with $\beta$ 
 velocity laws, for  $\beta\,>\,$1.
These observational discrepancies with CAK and m-CAK theories could be related  to  a  change in 
the  parameter $\alpha$ along the wind due either to a change in the ionization of the wind or to a decoupling of the line-driven ions in the wind from the ambient gas 
(\cite[Achmad et al. 1997]{Ach97}). 
In 2004, \cite[Cur\'e (2004)]{Cur04} revisited the theory of steady fast-rotating line-driven 
winds and found that for $\omega$=\,V/V$_{crit}$$>70$\% there exists another hydrodynamical or {\it slow} solution, 
which is denser and slower than the standard m-CAK solution. 
For B-type stars he obtained velocity distributions, as well as the critical point and $\dot{M}$, that
can be matched by a velocity law with $\beta\,>\,$1. Therefore, we propose here that the winds of Asg could be related to  the {\it slow hydrodynamical
solutions} described by \cite[Cur\'e (2004)]{Cur04}. We demonstrate numerically that for Asg, given a particular set of wind parameters,  
$k$, $\alpha$ and $\delta$, there exists a {\it slow 
solution} at all latitudes for low rotational velocities ($\omega$\,$<$\,0.4), and even for the non-rotating case. 
\firstsection
\section{Results and Conclusions}
\begin{figure}
\begin{center}{
 \includegraphics[width=1.85in]{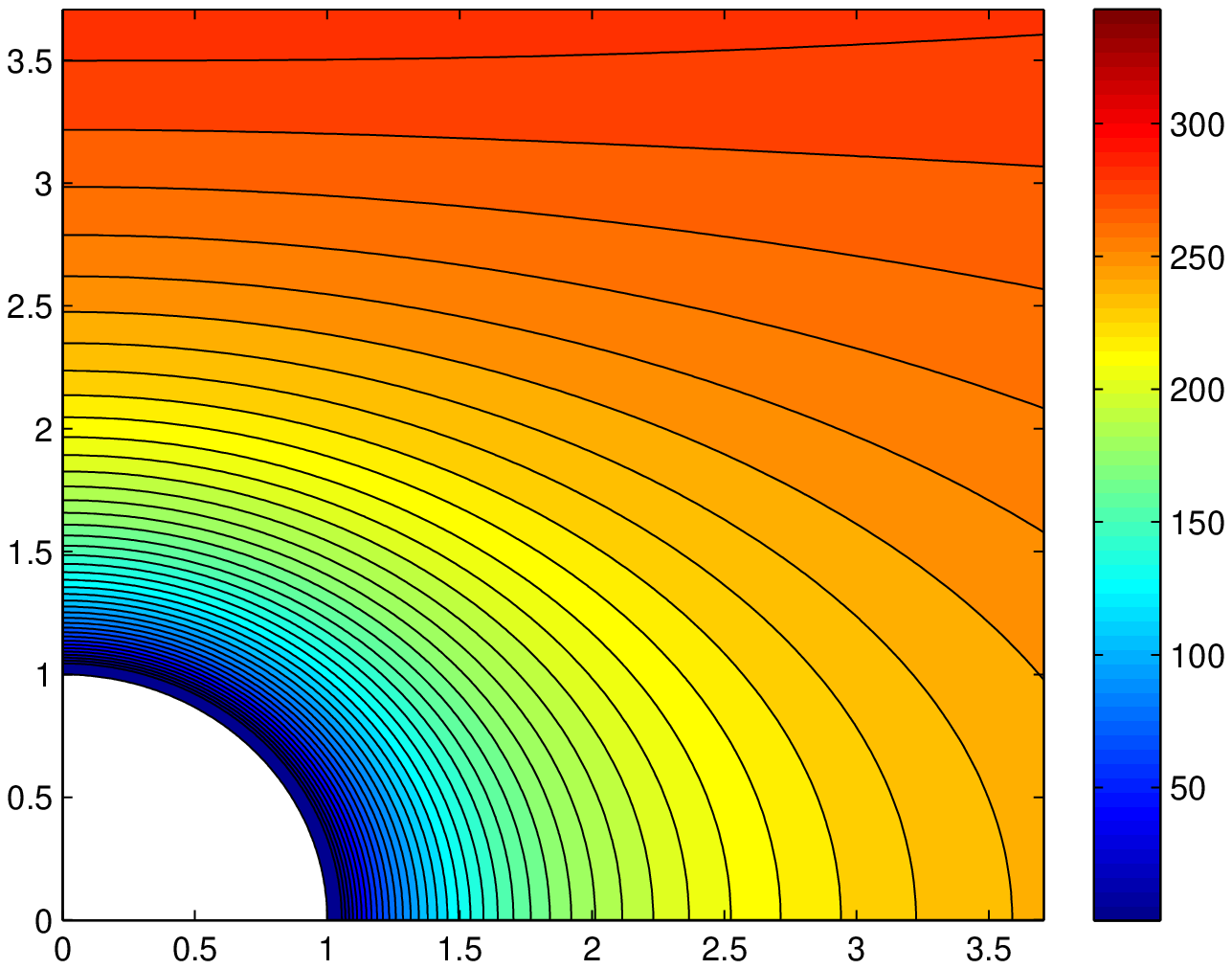} \includegraphics[width=1.85in]{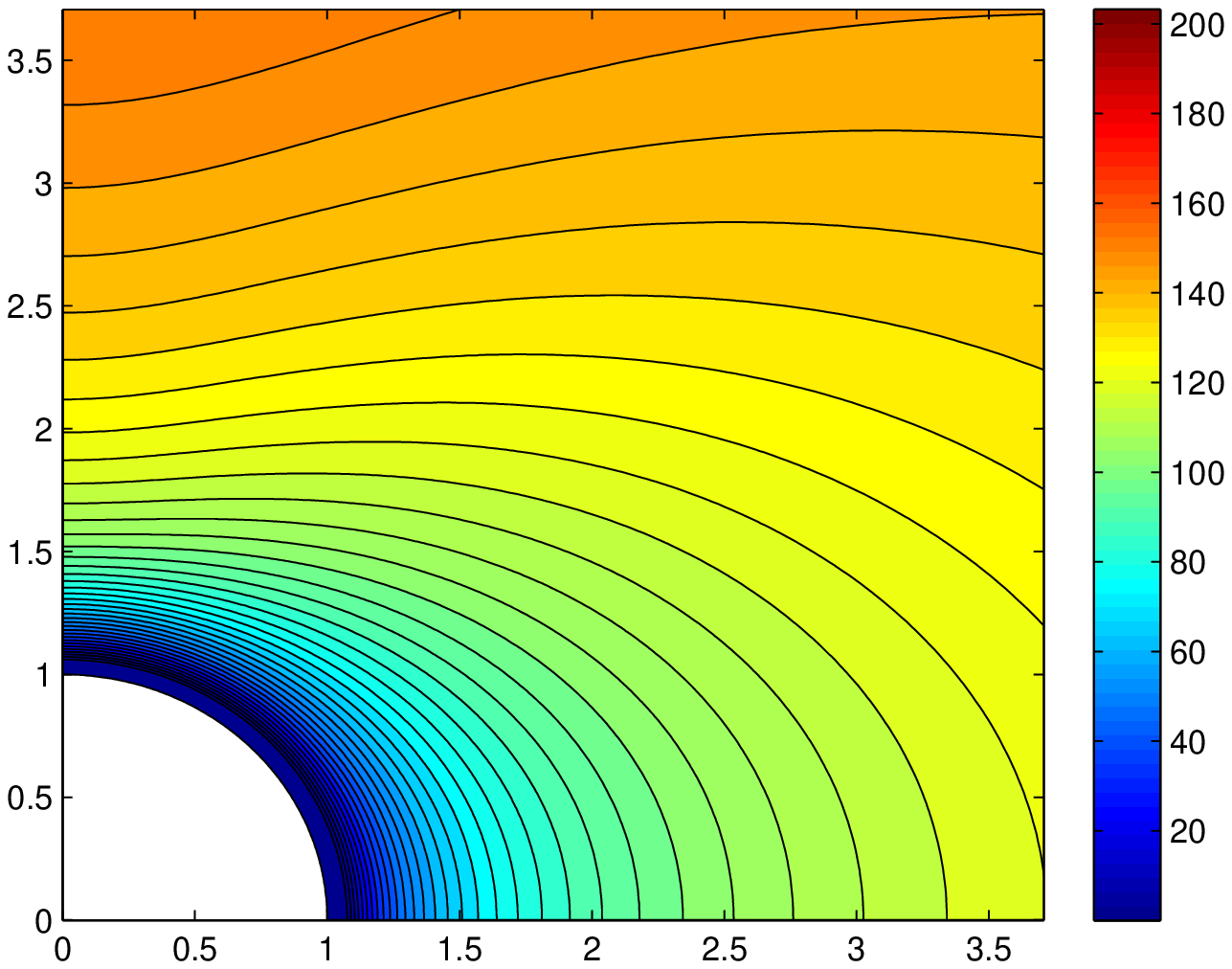}
 } 
 \caption{Wind velocity distributions. Left: Model\,2 (Fast). Right: Model\,3 (Slow). }
   \label{fig1}
\end{center}
\end{figure}

\begin{figure}[h]
\begin{center}{ \includegraphics[width=3.8in]{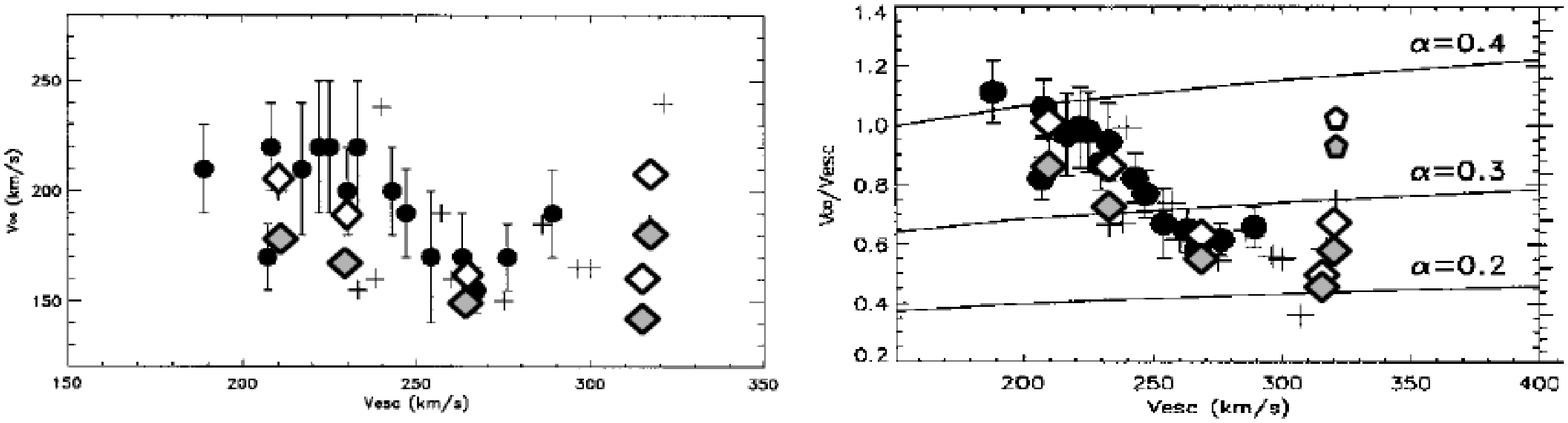} }
 \caption{The white/grey romboids (penthagons) correspond to equator/pole slow (fast) solutions, black symbols correspond to
 Figs. 2 and 3 of \cite[Verdugo et al. (1998)]{Ver99}}.
   \label{fig2}
\end{center}
\end{figure}

We calculated numerically wind solutions for different rotational velocities and different
 stellar and wind parameters. Table \ref{tab1} lists some of the models we
 computed for low values of $\omega$:
  those marked with an "s" in the first column present a slow solution at all
  latitudes, while the remaining corresponds to the 
   fast solution at all latitudes ("f"). Figure\,\ref{fig1} displays the behaviour of the fast 
   and slow solutions as a function of latitude for models with the same stellar and wind 
   parameters, except $\delta$. $\dot{M}$ and V$_{\infty}$ obtained with models 1, 3, 4, 5 and 
6 are in good agreement with those observed in Asg (\cite[Verdugo et al. 1999]{Ver99}; \cite[Kudritzki et al. 1999]{Kud99}). 
Instead, the fast solution leads to higher $\dot{M}$
 and V$_{\infty}$. We find that  the solutions for slow winds (see 
 Figure\,\ref{fig2}, in white/grey romboids) match observational data of some Asg stars
  (black symbols, \cite[Verdugo et al. 1998]{Ver99}).

  By solving the 1-D hydrodynamic equation of rotating line-driven winds of 
  Asg  for different sets of line force parameters, we found 
  a slow wind regime over all latitudes when increasing the parameter $\delta$, which  
  describes changes in the ionization stage of the wind. 
These slow solutions trace the observational trends found by 
\cite[Verdugo et al. (1999)]{Ver99}. 
Our result supports these authors'  
hypothesis, stating that the negative slope of 
V$_{\infty}$/V$_{esc}$\,vs.\,V$_{esc}$ could be
related to the degree of ionization and wind density.

\begin{table}
  \begin{center}
  \caption{Stellar and Wind Parameters}
 \label{tab1}
\firstsection 
 {\scriptsize
  \begin{tabular}{lrcccccccccccc}
        \hline
Mod&Teff&log\,g&R&$\omega$&$\alpha$&$k$&$\delta$&$\alpha_{eff}$&V$\infty_{pol}$&F$_{m, pol}$&V$\infty_{eq}$&F$_{m, eq}$&$\dot{M}$\\
&\tiny{[K]}&&\tiny{[R$_{\odot}]$}&&&&&&\tiny{[km\,s$^{-1}$]}&\tiny{[M$_{\odot}$/(yr\,sr)]}&\tiny{[km\,s$^{-1}$]}&\tiny{[M$_{\odot}$/(yr\,sr)]}&\tiny{[M$_{\odot}$/yr]}\\\hline
1 (s)&13000&1.73&68&0.4&0.51&0.03&0.23&0.28&160&4.78$\times\,10^{-5}$&142&5.78$\times\,10^{-5}$&6.41$\times\,10^{-4}$\\
2 (f)&10000&2.0&60&0.4&0.49&0.07&0.15&0.34&350&1.14$\times\,10^{-5}$&291&1.54$\times\,10^{-5}$&1.59$\times\,10^{-4}$\\
3 (s)&10000&2.0&60&0.4&0.49&0.07&0.26&0.23&207&9.79$\times\,10^{-8}$&181&1.21$\times\,10^{-7}$&1.32$\times\,10^{-6}$\\
4 (s)&9500&1.7&80&0.4&0.49&0.07&0.26&0.23&168&5.94$\times\,10^{-7}$&149&7.40$\times\,10^{-7}$&8.04$\times\,10^{-6}$\\
5 (s)&9000&1.7&100&0.4&0.49&0.07&0.26&0.23&188&3.08$\times\,10^{-7}$&165&3.84$\times\,10^{-7}$&4.18$\times\,10^{-6}$\\
6 (s)&9000&1.7&120&0.4&0.49&0.07&0.26&0.23&206&3.57$\times\,10^{-7}$&178&4.49$\times\,10^{-7}$&4.87$\times\,10^{-6}$\\\hline
  \end{tabular}
  }
 \end{center}
\end{table}

\vspace*{-0.3 cm}
\firstsection

\end{document}